\documentstyle[12pt]{article}
\def\degr{\hbox{$^\circ$}}
\def\arcmin{\hbox{$^\prime$}}
\def\arcsec{\hbox{$^{\prime\prime}$}}
\def\utw{\smash{\rlap{\lower5pt\hbox{$\sim$}}}}
\def\udtw{\smash{\rlap{\lower6pt\hbox{$\approx$}}}}

\def\farcm{\hbox{$.\mkern-4mu^\prime$}}
\def\farcs{\hbox{$.\!\!^{\prime\prime}$}}

\def\a{{$\alpha$}}

\def\g67{{G~67.7+1.8}}
\def\315{{G~31.5--0.6}}
\def\51c{{G~49.2--0.7}}
\newcommand{\h}{$^{\rm h}$}
\newcommand{\m}{$^{\rm m}$}
\newcommand{\s}{$^{\rm s}$}
\newcommand{\dd}{$\delta$}
\newcommand{\ha}{\rm H$\alpha$}
\newcommand{\hbeta}{\rm H$\beta$}
\newcommand{\HII}{\ion{H}{ii}}
\newcommand{\hnii}{{\rm H}$\alpha+[$\ion{N}{ii}$]$}
\newcommand{\nii}{$[$\ion{N}{ii}$]$}
\newcommand{\sii}{$[$\ion{S}{ii}$]$}
\newcommand{\oii}{$[$\ion{O}{ii}$]$}
\newcommand{\oiii}{$[$\ion{O}{iii}$]$}
\newcommand{\snr}{\rm supernova remnant}
\newcommand{\snrs}{\rm supernova remnants}
\newcommand{\et}{et al.}
\newcommand{\flux}{$10^{-17}$ erg s$^{-1}$ cm$^{-2}$ arcsec$^{-2}$}
\newcommand{\dens}{\rm cm$^{-3}$}
\DeclareMathAlphabet{\mathsc}{OT1}{cmr}{m}{sc}
\def\testbx{bx}%
\DeclareRobustCommand{\ion}[2]{%
\relax\ifmmode
\ifx\testbx\f@series
{\mathbf{#1\,\mathsc{#2}}}\else
{\mathrm{#1\,\mathsc{#2}}}\fi
\else\textup{#1\,{\mdseries\textsc{#2}}}%
\fi}
\begin{document}

\title{The supernova remnants G67.7+1.8, G31.5--0.6 and G49.2--0.7}

\author{F. Mavromatakis$^1$, J. Papamastorakis$^{1,2}$, J. Ventura$^{1,2,3}$, 
W. Becker$^3$,  E. V. Paleologou$^{2}$,D. Schaudel$^3$}

   \maketitle
University of Crete, Physics Department, P.O. Box 2208, 710 03 Heraklion, Crete, Greece 

Foundation for Research and Technology-Hellas, P.O. Box 1527, 711 10 Heraklion, Crete, Greece

Max-Planck Institut f\"ur extraterrestrische Physik, Giessenbachstrasse, D-85740 Garching, Germany

   \begin{abstract}
Optical CCD imaging and spectroscopic observations of three supernova 
remnants have been performed for the first time. Filamentary and 
diffuse emission is discovered from the \snr\ \g67\ located $\sim$ 82\arcmin\  
to the south of CTB 80's pulsar. 
The \ha\ and sulfur emission are almost equally strong at 
a level of $\sim$ 20 $\times$ \flux\ suggesting shock heated emission. 
Electron densities less than 240 \dens\ are estimated, while   
the weak \oiii\ emission suggests shock velocities in the range of 
60--80 km s$^{-1}$. 
Emission can also be seen in the ROSAT All
Sky Survey data which indicate an extended hard X-ray source. 
Emission from \315\ is detected only in the \hnii\ image 
at a typical flux level of 35 $\times$ \flux. The morphology of the 
observed radiation is diffuse and partially correlated with the 
non--thermal radio emission. Deep long slit spectra 
detect sulfur line emission which is not strong enough to identify it 
as emission from shocked gas.
Finally, optical emission from \51c\ is obscured by several 
dark nebulae which probably give rise to significant X-ray 
attenuation. 
The \hnii\ flux is typically $\sim$ 40 $\times$ \flux\ while the \sii\ 
flux is very weak, not allowing its identification as shock heated. 
However, a small area of $\sim$ 3\arcmin\ $\times$ 1\arcmin\ emits strong 
sulfur flux relative to \ha\ (\sii /\ha\ $\sim$ 0.6). 
This area is located in the south--east of 
\51c, close to the outer boundaries of the X--ray and radio emission.
However, deep optical spectra would be required to firmly establish 
the nature of this emission and its association to \51c.

   \end{abstract}
\section{Introduction}
Most supernova remnants have been discovered by their non--thermal 
synchrotron radio emission and their shell morphology. 
Optical observations may detect light from a remnant 
depending on the distance and age of the remnant, and the properties 
of the local interstellar medium. 
The interstellar medium is not homogeneous or uniform,
encompassing denser regions of interstellar ``clouds''. It
is the interaction of these clouds with the primary
shock wave of a middle aged remnant that ultimately gives 
rise to optical radiation.
Imaging observations of \snrs\ use interference filters to isolate main 
optical emission lines like \ha\ 6563 \AA, \hbeta\ 4861 \AA, \sii\ 
6716, 6731 \AA\ and \oiii\ 5007 \AA. The \sii\ to \ha\ ratio 
serves as a discriminator between \HII\  and shock heated emission, 
although in limiting cases supplementary data on the target ought be sought 
(e.g. Fesen \et\ ~\cite{fes85}). 
Information about the amount of interstellar extinction can be extracted from
the \ha\ to \hbeta\ ratio while possible variations of this ratio over the  
remnant's extent may indicate interaction with the local 
interstellar medium (e.g. Osterbrock \cite{ost89}). 
Provided that the \oiii 5007 \AA\ line 
is observed in a remnant, its intensity relative to \hbeta\ can provide 
valuable information about the shock speed 
(e.g. Cox \& Raymond ~\cite{cox85}). 
Spectroscopic observations on the other hand offer the advantage of more 
detailed spectral information allowing for comparison with published 
shock models, but at the expense of limited spatial coverage.  
\par
In an effort to broaden our knowledge on the least observed supernova 
remnants, we performed optical observations of three known radio remnants 
that had not been detected before in optical wavelengths.
The supernova remnant G 67.7+1.8 was first detected in a galactic 
plane radio survey at 327 MHz by Taylor \et\ (\cite{tay92}) using the 
Westerbork Synthesis Radio Telescope. The authors proposed its identification 
as a supernova remnant based on its dual--arc morphology and spectral index 
of \a\ $=$ --0.5 (S$_{\nu}$ $\sim$ $\nu$$^{\alpha}$). It is characterized by 
an angular diameter of $\sim$ 9\arcmin\ and a flux at 1 GHz of 
$\sim$ 1.2 $\times$ 10$^{-21}$ W m$^{-2}$ Hz$^{-1}$ sr$^{-1}$ (Taylor 
\et\ \cite{tay92}). A search in the literature for references to X-ray or 
optical observations turned out negative (Neckel \& Vehrenberg \cite{nec87}).
However, a careful examination 
of the red POSS plates reveals faint but filamentary emission along the 
north part of the shell of \g67, while diffuse X--ray emission is also seen 
in the ROSAT All Sky Survey data. 
\par
The shell--like morphology of the radio continuum emission at 4750 MHz 
of \315\ and the non--thermal emission led F\"urst \et\ (\cite{fur87}) 
to propose the identification of this object as a supernova remnant. 
The spectral index is found in the range of 
--0.2 to --0.5 with a flux density of $\sim$ 1.8 Jy at 4750 MHz. 
The POSS plates 
do not show any traces of optical emission that could be attributed to 
G 31.5--0.6,   
 while the detection of X-ray emission is not reported in the literature. 
Case and Bhattacharya (\cite{cas98}) proposed a distance of 16.7 kpc to 
\g67\ based on the $\Sigma$ -- D relation and a distance of 12.9 kpc to 
\315. 
\par
The third target of our observations was the supernova remnant 
G 49.2--0.7. 
This remnant is also known as W51C because it belongs to the radio 
complex W51, including the \HII\ regions W51A and W51B. The 330 MHz radio image 
of Subrahmanyan and Goss (\cite{sub95}) shows an extended structure of 
angular dimensions $\sim$ 50\arcmin\ $\times$ 35\arcmin\ 
while the ROSAT soft X--ray data also suggest a similar extend (Koo \et\ 
\cite{koo95}). The spectral analysis of the ROSAT data showed that a thermal 
model could account for the observed spectrum. Koo \et\ (\cite{koo95})  quote 
a shock temperature of $\sim$ 3 $\times$ 10$^6$ K, a shock velocity of 
$\sim$ 500 km s$^{-1}$ and an age of $\sim$ 30000 yrs. The authors proposed
that the remnant is located $\sim$ 6 kpc away. 
\par
In this work we present deep CCD images of the forementioned 
remnants in \hnii, \sii, \oii\ and \oiii.
Information about the observations and the data 
reduction is given in Sect. 2. In Sect. 3, 4 and 5 we present the results of 
our imaging observations. We also discuss the results from
the long slit spectra taken at specific locations of interest. 
Finally, in Sect. 6 we discuss the physical properties of \g67\ and 
the implications of the current observations  to the properties of 
\315\ and \51c. 
\section{Observations}
\subsection{Optical images}
The observations presented here were performed with the 0.3 m
telescope at Skinakas Observatory. The fields of the radio remnants 
were observed in June 16, and July 08--11, 1999. 
Two different CCDs were used during the observations.   
The first, was a 1024 $\times$ 1024 Thomson CCD 
which resulted in a  69\arcmin\ $\times$ 69\arcmin\ field of view 
and an image scale of  4\arcsec\ per pixel. 
The second was a 1024 $\times$ 1024 Site CCD which had a larger pixel size 
resulting in a 89\arcmin\ $\times$ 89\arcmin\ field of view and an image 
scale of 5\arcsec\ per pixel. 
The characteristics of the interference filters are listed in 
Table~\ref{filters} while   
the number of frames taken in each filter is given in Table~\ref{obs}. 
The exposure time of 
a single frame is 1800 s. The final images in each filter are the average 
of the individual frames. 
All coordinates quoted in this work refer to epoch 2000.
\g67\ was also observed with the 1.3 m telescope at Skinakas 
Observatory on August 21, 2000. The object was imaged with the 
\hnii\ filter, is not flux calibrated and is characterized by a scale 
of 1\arcsec\ per pixel.
\par
Standard IRAF and MIDAS routines were used for the reduction of the data. 
Individual frames were bias subtracted and flat-field corrected using 
well exposed twilight flat-fields. The spectrophotometric standard stars 
HR7596, HR7950, HR8634, and HR718 were used for flux calibration.
\subsection{Optical spectra}
Long slit spectra were obtained on August 22 and 23, 2000 using the 
1.3 m Ritchey--Cretien telescope at Skinakas Observatory. 
One long slit spectrum of \g67\ was obtained on July 18, 1999 and is 
flux calibrated. All other spectra are not flux calibrated. 
The spectrophotometric standard stars HR718 and HR7596 were used to 
determine the detector's sensitivity function.    
The data were taken with a 1300 line mm$^{-1}$ grating 
and a 800 $\times$ 2000 Site CCD having a 
15 $\mu$m pixel size which resulted in a 1.04 \AA\ pixel$^{-1}$. 
The slit had a width of 7\farcs7 and, in all cases, was oriented
in the south-north direction.  
The number of available spectra from each remnant and the exposure time 
of each spectrum are given in Table ~\ref{spectra}. 
%
\section{The supernova remnant G 67.7+1.8}
\subsection{The \hnii\ and \sii\ line emission}
The \snr\ \g67\ appears as a $\sim$ 9\arcmin\ long and 
$\sim$ 10\arcsec\ wide filament in the \hnii\ and \sii\ images, oriented in 
the SW to the NE direction (Fig. \ref{fig01}). 
Diffuse emission to the south of the filament is present but still 
within the boundaries of the radio emission. 
In Table ~\ref{fluxes} we list typical fluxes measured in the calibrated 
images of the observed remnants. In cases where we failed to detect 
emission in a specific filter, the 3$\sigma$ upper limit is quoted.  
The \hnii\ image also shows a small scale, diffuse
structure $\sim$ 12\farcm3 to the SW of the \g67, at 
\a\ $\simeq$ 19\h53\m57\s\ and \dd\ $\simeq$ 31\degr21\arcmin54\arcsec\
(Fig. \ref{fig02}). 
This structure has a typical extent of $\sim$ 3\farcm3 and  
emits \hnii\ radiation at a level of $\sim$ 20--35 $\times$ \flux. 
It is probably unrelated to \g67\ since the detected emission is well 
outside the faintest radio contours. 
A single 2400 s long slit spectrum was obtained from this object, 
and the results are given in Table ~\ref{sfluxes}.  
We designate this source as GAL 67.58$+$1.88 since it was not 
previously catalogued.
\subsection{The \oiii\ and \oii\ images}
The filament first seen in \hnii\ is also detected in the oxygen forbidden 
lines of 5007 \AA\ and 3727 \AA\ (images not shown here) 
but the emission is quite weak. 
Continuum subtraction and a light smoothing on the resulting images 
were necessary in order to clearly identify the filament. 
The overall length of the filament 
in \oiii\ is $\sim$ 8\arcmin, however, the emission is not spatially continuous 
but two gaps are present. Interestingly, GAL 67.58+1.88 is also seen in 
these filters. It possesses a patchy 
appearance in the \oiii\ filter while the \oii\ emission looks like an arc 
convex to the south. We estimate an \oiii\ flux of $\sim$ 3 $\times$ \flux\ and
an \oii\ flux of $\sim$ 5 $\times$ \flux.
%
\subsection{The G 67.7+1.8 low resolution spectrum}
The spectrum taken from \g67\ (Table ~\ref{sfluxes}) 
shows that we observe optical 
radiation originating from shocked gas since we estimate 
\sii/\ha\ $\sim$ 1.2 ($\pm$ 0.1) 
and the optical filament is well correlated 
with the 1400 MHz and 4850 MHz radio data (Condon \et\ \cite{con94}, 
Fig. \ref{fig02}). 
The sulfur line ratio of 1.28 ($\pm$ 0.08) suggests a low electron density 
$\sim$ 142 \dens, though taking the statistical error
into account implies that
densities in the range of 60 -- 240 \dens\ would be compatible with our
measurement. 
Finally, the \hbeta\ flux is rather low compared to the \ha\ flux 
suggesting significant interstellar extinction 
(\ha/\hbeta\ $\sim$ 10.9 $\pm$ 2.6). 
\par
The spectrum of GAL 67.58$+$1.88 does not allow a reliable 
determination of the nature of this object due to the large errors in 
the sulfur lines. However, the strong \nii\ lines would suggest 
a circumstellar origin of the extended emission.
\subsection{The ROSAT All Sky Survey data}
In the course of the ROSAT all-sky survey, \g67\ was in the PSPC field of 
view between Oct 22-25, 1990 for a total exposure time of $\sim 530$ sec.
About 73 events were detected above the background level, 
in a circular area of 8\arcmin\ radius, at energies higher than 0.5 keV. 
No emission is seen above the background below 0.5 keV. 
The counts detected above 0.5 keV imply a surface brightness 
of $\sim 6.9 (\pm1.4) \times 10^{-4}$ cts s$^{-1}$ arcmin$^{-2}$. 
According to Dickey \& Lockman (1990), the galactic absorption along the 
line of sight is $10^{22}$ cm$^{-2}$.  
Assuming a thermal bremsstrahlung spectrum and fixing N$_{\rm H}$
to values in the range of $0.5 -- 1.0 \times 10^{22}$ cm$^{-2}$, 
we find temperatures 
of 0.2--0.3 keV, equivalent to blast wave speeds in the range of 
$\sim$ 400 -- 500 km s$^{-1}$. A thermal blackbody spectrum requires 
lower temperatures of the order of $\sim$ 0.15 keV. The low number of the  
detectedwe photons do not allow us to uniquely, identify the nature of X--ray 
emission. 
According to the NVSS data, three faint radio sources show up close 
to the center of \g67 which could be indicative of emission from a young 
neutron star. 
Lorimer et al. (\cite{lor98}) searched for radio emission from a 
pulsar in the area of \g67 using the Jodrell Bank Radio facility.
No radio pulsar was detected down to a level of 0.8 mJy.
\section{The supernova remnant G 31.5--0.6}
The radio contours at 4850 MHz  (Condon \et\ \cite{con94}), plotted 
linearly from 0.02 Jy/beam to 0.30 Jy/beam, are overlaid to our 
\hnii\ image (Fig. \ref{fig03}). The correlation of the optical and 
radio data may suggest their physical association, although 
the lack of strong \sii\ emission makes this identification 
very difficult. The observed optical emission appears as a broad 
incomplete shell of diffuse emission convex to the NW.  
Typical \hnii\ fluxes and the 3$\sigma$ upper limits on the 
\sii, \oii\ and \oiii\ fluxes are given in Table ~\ref{fluxes}. 
%
\subsection{The spectrum of G 31.5--0.6}
The analysis of the optical images showed that strong \hnii\ 
emission is present to the south--east and north--west areas.
However, the latter area of emission coincides with the location 
of an elongated small diameter source (GAL 31.650--00.649) 
reported by  F\"urst \et\ (~\cite{fur87}) 
which is characterized by a flat radio spectrum. Consequently, the slit was 
placed at the former area where non--thermal emission was detected and 
at a right ascension of 18\h51\m49\s\ and a declination of -1\degr34\arcmin
20\arcsec. 
The spectra taken in this area 
show that the \ha\ emission is stronger than the sulfur emission (Table 
~\ref{sfluxes}).  
\section{The supernova remnant G 49.2--0.7}
The supernova remnant \51c\ lies close to the galactic plane 
along with several \HII\ regions as well as with dark nebulae being 
projected on it (Fig. ~\ref{fig04}). A search in the SIMBAD database 
resulted in $\sim$ 25 \HII\ regions within a circular field of 1\degr\ 
diameter. However, only three \HII\ regions overlap W51C and these are 
GAL 049.2-00.7, GAL 049.0-00.6 (Wilson \et\ ~\cite{wil70}) and SH 2--79 
(Acker \et\ ~\cite{ack83}). 
The observed optical emission occupies an angular extent of 
$\sim$ 40\arcmin\ $\times$ 40\arcmin\ and 
the morphology in the \hnii\ and \sii\ filters is diffuse. The observed 
radiation seems to split into two parts separated by a dark lane of material 
running along \a\ $\simeq$ 19\h22\m50\s. 
The east part shows several patches of emission in \hnii\ while the west 
part appears more diffuse. 
The sulfur line flux is relatively weak and thus, the image is not shown here.  
Optical diffuse or filamentary emission from the \51c\ area is not detected 
in our \oii\ and \oiii\ images (Table ~\ref{fluxes}).
The diffuse emission observed to the north--west  
of the dark lane may be associated to W51B. 
The optical emission west of this lane and south of 13\degr55\arcmin\ is 
probably not related to W51B but even its relation to W51C is not clear 
since it is located outside the main body of the radio emission of W51C.  
However, some radio contours at 330 MHz (Subrahmanyan and Goss \cite{sub95}) 
do overlap this optical emission. 
\subsection{The optical spectrum of G 49.2--0.7}
The slit was placed at a bright spot of  the diffuse emission seen in 
the east, in the \hnii\ image, which coincides with the area of radio 
emission from the remnant W51C. The area west of $\sim$ 19\h23\m\ is 
mainly dominated by W51B which is an \HII\ region. The slit was placed at 
\a\ $=$ 19\h23\m09\s\ and \dd\ $=$ 13\degr59\arcmin39\arcsec\ 
and the signal to noise weighted average fluxes of the detected lines 
are shown in Table ~\ref{sfluxes}, where it is seen that the sulfur emission is 
weak relative to the \ha. 
\section{Discussion}
The \snrs\ \g67\ and \315\ are among the least observed remnants both in 
radio and optical wavelengths. This is not true for \51c\ where   
extended radio and X--ray observations have revealed its physical properties.   
\subsection{The \g67\ radio remnant}
The radio remnant \g67\ is detected 
for the first time in the optical band as well as in the soft X--ray band by 
ROSAT. Both the positional correlation 
and the nature of the optical spectrum provide convincing evidence that the 
observed emission is indeed associated to \g67. The long slit spectra 
suggest a low electron density ($\sim$ 140 \dens) but even the small (6\%) 
error on the sulfur line ratio cannot exclude densities in the range 
of 60--240 \dens. The shock velocity is estimated to be less than 
100 km s$^{-1}$  given the weak \oiii\ emission and probably will lie in the 
range of 60--80 km s$^{-1}$ (Cox \& Raymond ~\cite{cox85}, Hartigan \et\ 
~\cite{har87}) while the strong sulfur emission relative to \ha\ suggests 
a partially neutral medium. In order to obtain a better insight to the 
properties of \g67\ a reliable distance determination is necessary. 
However, given the limited number of available observations, the $\Sigma$ -- D 
relation is the only tool available for this purpose. Case \& Bhattacharya 
(~\cite{cas98}) quote a distance of 16.7 kpc but the large errors on the 
proportionality factor and the exponent of the $\Sigma$ -- D relation allow 
a wide range of distances from $\sim$ 7 -- 27 kpc. 
In view of the optical observations reported here, distances less than 
$\sim$ 17 kpc are more probable, otherwise detection of optical radiation 
would be very difficult due to interstellar extinction. The 
\ha /\hbeta\ ratio of $\sim$ 11 corresponds to an interstellar 
extinction of 1.7 ($\pm$0.3) and thus, supports distances much lower than 
17 kpc (Hakkila \et\ \cite{hak97}). Another approach to a distance estimate 
would involve the 
measured electron density and assumptions about the shock velocity and 
initial explosion energy. 
In the following we will assume a shock velocity V$_{\rm s}$ 
of 70 km s$^{-1}$, and an 
explosion energy E in the range of 10$^{50}$ --  10$^{51}$ ergs. An energy of 
10$^{51}$ ergs is considered as the typical energy released in a supernova 
explosion. With the aid of the relation 
\begin{equation}
{\rm n_{[SII]} \simeq\ 45\ n_c \times (V_s/100\ km\ s^{-1})^2}
\end{equation}
given by Fesen \& Kirshner (\cite{fes80}), the above assumptions and the 
measured 
range of electron densities, we estimate that the preshock cloud densities 
${\rm n_c}$ will lie in the range of 3 -- 11 \dens.
McKee \& Cowie (~\cite{mck75}) derived an equation relating the energy of 
explosion, shock radius and shocked cloud parameters as 
\begin{equation}
{\rm E} = 2 \times 10^{46} \beta^{-1} 
{\rm n_c}\ (V_{\rm s}/100\ {\rm km\ s}^{-1})^2 \ 
({\rm r_s/1\ pc})^3 \ \ {\rm erg,} 
\end{equation}
where $\beta$ is a factor of the 
order of 1--2, and ${\rm r_s}$ is the shock radius in pc.  
Using Eq. (2) and the range of preshock cloud densities, 
we find that for an explosion energy of  10$^{50}$ erg the distance 
should lie in the range 7 -- 12 kpc, while for an explosion energy 
of 10$^{51}$ erg the distance should lie in the range of 16 -- 26 kpc. 
These naive calculations show that the energy released during the 
supernova explosion must be significantly less than the canonical energy 
of 10$^{51}$ erg. The soft X--ray data are equally well fitted by 
thermal bremsstrahlung or blackbody models and suggest temperatures in 
the range of 0.15 -- 0.3 keV. However, the low counting statistics do not 
allow for a reliable determination of the column density, temperature and 
X--ray flux. Dedicated X--ray observations would  be 
required to better understand the physical properties of \g67\ and its 
surroundings.  
\par
A new source of diffuse emission is detected close to \g67, though not 
related to the remnant. The current observations suggest strong 
\oiii, \nii, \ha\ but weak \sii\ emission characteristic of emission of
circumstellar origin. The bright blue star GSC 02669--04343 is found at 
the west boundary of GAL 67.58$+$1.88 but it is not clear if they are 
related in any way. Higher resolution imaging and spectral observations 
would be needed to study this source in detail.
\subsection{\315, a source of weak sulfur line emission}
A search in the SIMBAD database 
revealed several \HII\ regions and dark nebulae within the field of \315.
The candidate supernova remnant GAL 31.7--1.0 (Gorham ~\cite{gor90}) 
is located to the south--east of \315\ but 
we do not find any strong signs of optical emission. The \hnii\ emission from 
the vicinity of \315\ is diffuse with the \HII\ region GAL 31.65--00.649 
superposed on its north--west part. The positional correlation between the 
optical and radio data and their similar shapes would suggest the 
identification of the optical flux as emission from \315, even though 
a chance superposition can not be excluded. 
The lower limit on the \ha/\hbeta\ ratio ($>$ 12) translates to a lower limit 
on the neutral hydrogen 
column density of 8 $\times$ 10$^{21}$ cm$^{-2}$ which is consistent 
with the column density of $\sim$ 1.3 $\times$ 10$^{22}$ cm$^{-2}$ given 
by Dickey \& Lockman (~\cite{dic90}). 
The ratio of the sulfur lines to \ha\ is 0.27 which is lower than the limit of 
$\sim$ 0.4 required to optically identify a supernova remnant. 
Although the fluxes of the sulfur lines are relatively accurately established 
(7--8$\sigma$), their line ratio is consistent with electron densities lower 
than $\sim$ 380 \dens. 
The low value of the \sii\ / \ha\ ratio and the high value of the 
F(6716)/F(6731) ratio are more suggestive of a spectrum of an \HII\ region
rather than of a SNR spectrum. Thus, the current data cannot identify the 
observed emission as emission from shocked gas, despite the positional 
correlation.
\subsection{The complex around \51c}
Extended optical emission is present in the \hnii\ filter from \51c, 
while it is substantially reduced in the \sii\ filter. 
No optical emission, at our sensitivity threshold, is 
detected in both oxygen filters. 
Long slit spectra obtained from the brighter areas seen in the 
\hnii\ filter do not suggest emission from shock heated material. 
Both the 
\sii/\ha\ and the sulfur lines ratio are indicative of \HII\ emission 
(e.g. Fesen \& Hurford ~\cite{fes95}). It is possible that the emission 
within the slit was dominated by the \HII\ region G 049.0--00.6 detected 
in an H 109$\alpha$ survey by Wilson \et\ (\cite{wil70}), even though 
the authors state that due to the complexity of the region it is difficult 
to accurately measure the sizes and temperatures of these regions. 
Koo \et\ (\cite{koo95}) have presented evidence for fast moving molecular 
gas which blocks our view to the west areas of W51 and may be responsible
for the dark lane present in the optical data as well as in the X-ray data. 
The authors also found variations in the column density across the source 
of X-ray emission while the gas temperature remained essentially constant. 
Even though, the \sii\ emission is generally weak, a 
$\sim$ 3\arcmin\ $\times$ 1\arcmin\ area seems to emit 
stronger sulfur flux at a level of $\sim$ 5 $\times$ \flux.
The corresponding \hnii\ flux suggests that we may be observing 
emission from shocked gas since we estimate a \sii/\ha\ ratio 
of $\sim$ 0.6. This region is located 
in the south--east boundary of the X-ray and radio emission and 
specifically, at a location of strong soft X--ray emission (Fig. 3a of 
Koo \et\ ~\cite{koo95}). The presence of \HII\ regions, variable X-ray 
attenuation and molecular flows may render impossible the detection 
of shock heated emission from \51c. Nevertheless, it is possible 
that certain areas in the south of \51c\ suffer less absorption and 
some optical emission may escape the remnant unobscured. Long slit 
spectra at the specified location should be able to determine 
unambiguously whether 
the detected emission is shock heated or not. 
\section{Conclusions}
Three, not so well known, supernova remnants were observed and 
detected for the first time in the optical band.
A thin long filament is detected in the north boundary of the 
radio emission from \g67. Its spatial correlation to the radio emission  
and the long slit spectra suggest its identification as optical 
emission from a supernova remnant.  
A new faint structure called  GAL 67.58$+$1.88 is detected to the 
south--west of \g67\ but its nature is not clear yet. 
The imaging observations of \315\ detect \hnii\ emission which is found to be 
partially correlated with the radio emission. However, long slit spectra 
show that the sulfur emission is not strong enough 
to justify shock heated emission.  
Optical emission is detected from the area of \51c\ 
in the \hnii\ filter while the measured fluxes in the 
\sii\ filter are quite weak. 
A patch of $\sim$ 3\arcmin\ $\times$ 1\arcmin\ 
in the south--east emits more \sii\ flux than its surroundings. 
It could be possible that the south areas of 
\51c\ suffer less absorption, allowing for the detection of shock heated 
emission. However, the nature of the emitted radiation in the south--east
would be uniquely identified only through deep, long slit spectra.
\vfill\eject
  \begin{table}
      \caption[]{Interference filter characteristics}
         \label{filters}
\begin{flushleft}
\begin{tabular}{lllll}
            \hline
            \noalign{\smallskip}
Filter     & Wavelength$^{\rm a}$	& Line (\%)	\cr
           & (FWHM)(\AA)& contributions       \cr
            \hline
H\a\ +[NII] & 6555 (75)	& 100, 100, 100$^{\rm b}$	\cr
            \hline
[SII]       & 6708 (27)	& 100, 18$^{\rm c}$	\cr
            \hline
[OII]      & 3727 (28)	& 100, 100$^{\rm d}$		\cr
	    \hline
[OIII]      & 5005 (28)	& 100		\cr
            \hline
Cont red    & 6096 (134)&  --      \cr
		\hline
Cont blue   & 5470 (230)  & --     \cr
		\hline
\end{tabular}
\end{flushleft}
${\rm ^a}$ Wavelength at peak transmission for $f$/3.2\\\
${\rm ^b}$ Contributions from $\lambda$6548, 6563, 6584 \AA \\\
${\rm ^c}$ Contributions from $\lambda$6716, 6731 \AA \\\
${\rm ^d}$ Contributions from $\lambda$3727, 3729 \AA 
   \end{table}
  \begin{table}
      \caption[]{Log of the exposure times}
         \label{obs}
\begin{flushleft}
\begin{tabular}{lllllll}
            \noalign{\smallskip}
\hline
 	&  \hnii\	& \sii\		& \oiii\ & \oii\    \cr
\hline
G 31.5--0.6&5400$^{\rm a}$(3)$^{\rm b}$	&5400 (3)		&1800(1)	&1800(1)	 \cr
  \hline
G 49.2--0.7 &7200(4) 	&7200(4)		&1800(1)	&1800(1)	 \cr
  \hline
G 67.7+1.8 &3600(2)	&3600(2)		&3600(2)	&3600(2)	\cr
  \hline
\end{tabular}
\end{flushleft}
${\rm ^a}$ Total exposure times in sec \\\
${\rm ^b}$ Number of individual frames \\\
   \end{table}
  \begin{table}
      \caption[]{Spectral log}
         \label{spectra}
\begin{flushleft}
\begin{tabular}{lllllll}
            \noalign{\smallskip}
\hline
 \g67\ 	&	\315\	&	\51c\	&	GAL 67.58$+$1.88\cr
\hline
 3$^{\rm a}$	&	2	&	3	&	1 \cr
\hline
 3600$^{\rm b}$ s	&	2700 s	&	2400 s	&	2400 \cr
\hline
\end{tabular}
\end{flushleft}
${\rm ^a}$ Number of spectra collected \\\
${\rm ^b}$ Exposure time of individual spectra \\\
   \end{table}
  \begin{table}
      \caption[]{Typically measured fluxes}
         \label{fluxes}
\begin{flushleft}
\begin{tabular}{lllll}
            \hline
            \noalign{\smallskip}
 	& \g67\	&	\315\	&	\51c\	\cr	
\hline
\hnii\ 	& 40	&	35	&	30	\cr	
\hline
\sii\ 	& 20	&	$<$ 6	&	5	\cr	
\hline
\oiii\ 	& 2	&	$<$ 10	&	$<$ 5	\cr	
\hline
\oii\ 	& 3	&	$<$ 6	&	$<$ 6	\cr
\hline
\end{tabular}
\end{flushleft}
${\rm }$ Fluxes in units of \flux \\\
   \end{table}
  \begin{table*}
        \caption[]{Relative line fluxes}
         \label{sfluxes}
         \begin{flushleft}
         \begin{tabular}{lllllll}
     \hline
 \noalign{\smallskip}
                & \g67\ & {\rm G 67.58+1.88} & \315\ & \51c\ \cr
\hline
Line (\AA) & F$^{\rm a,b}$ & F$^{\rm a,b}$ & F$^{\rm a,b}$  &  F$^{\rm a,b}$ \cr
\hline
4861 \hbeta\   &  92 (24)$^{\rm c}$ & $<$ 516 & $<$ 85  &  $<$ 100       \cr
\hline
5007 [OIII]      & 89 (26) & 641 (25)    &  --  & --   \cr
\hline
6300 [O I]       & 283 (8) &  --   & --  &  --      \cr
\hline
6360 [O I]       &  98 (21) &  --   &  --  & --   \cr
\hline
6548 \nii\  	& 192 (11) & 777(21) & 144 (16)  & 118  (16)  \cr
\hline
6563 \ha\ 	& 1000 (3)& 1000 (17)& 1000 (3)   &  1000 (2) \cr
\hline
6584 \nii\  	& 631 (4)& 2368 (8)& 444 (25)   &   400 (5)\cr
\hline
6716 \sii\ 	& 647 (4)& 246 (59)& 159 (12)  &  137 (12)\cr
\hline
6731 \sii\ 	& 506 (5)& 144 (83)&  116 (15)    &   99 (18) \cr
\hline
\hline
\ha/\hbeta\ 	& 10.9 (24)& $>$ 1.94    & $>$ 11.8  &  $>$ 10    \cr
\hline
\sii/\ha\ 	& 1.15 (4) & 0.40 (50)       &  0.27 (10)  & 0.24 (10)    \cr
\hline 
F(6716)/F(6731)	& 1.3 (6) &  --    &  1.4 (19)    &  1.4 (22) \cr
\hline 
\end{tabular}
\end{flushleft}
 ${\rm ^a}$ Uncorrected for interstellar extinction 

${\rm ^b}$ Listed fluxes are a signal to noise weighted
average of the available spectra 

${\rm ^c}$ Numbers in parentheses 
represent the relative (\%) error of the quoted fluxes

${\rm }$ All fluxes normalized to F(\ha)=1000
\end{table*}
\vfill\eject
  \begin {figure}
    \caption{\g67\ imaged in the \hnii\ filter for 300 s with 
     the 1.3 m telescope. 
    The image has been smoothed to suppress the residuals 
     from the imperfect continuum subtraction.
     Here and in the following images north is up, east to the left and the 
     coordinates refer to epoch 2000.} 
     \label{fig01}
  \end{figure}
  \begin {figure}
    \caption{The radio 1400 MHz contours (Condon \et\ \cite{con94}) 
     of \g67\ overlaid to the \hnii\ image. 
     The radio contours scale from 8 $\times$ 10$^{-4}$ to 0.02 Jy/beam.  
     The image has been smoothed to suppress the residuals from the imperfect 
     continuum subtraction.} 
     \label{fig02}
  \end{figure}
  \begin {figure}
    \caption{The neighborhood around \315\ in the \hnii\ filter. 
     The original image is 70\arcmin\ $\times$ 70\arcmin\ and has been 
     smoothed to suppress the residuals from the imperfect 
     continuum subtraction. Shadings run linearly from 0.0 to 50 
      $\times$ \flux\ while the 4850 MHz radio contours (Condon \et\ 
      \cite{con94}) scale also linearly 
      from 0.02 Jy/beam to 0.30 Jy/beam.}
     \label{fig03}
  \end{figure}
  \begin {figure*}
    \caption{The supernova remnant \51c\ (W51C) imaged in the \hnii\ filter. 
     The image has been smoothed to suppress the residuals from the imperfect 
     continuum subtraction and the shadings run linearly from 0.0 to 50 
      $\times$ \flux. The arrow points to the area where enhanced \sii\ emission
      is detected.} 
     \label{fig04}
  \end{figure*}
\end{document}